\documentclass[a4paper,fleqn,usenatbib,useAMS]{mnras}
\usepackage{scalerel}
\usepackage{tikz}
\usetikzlibrary{svg.path}

\definecolor{orcidlogocol}{HTML}{A6CE39}
\tikzset{
 orcidlogo/.pic={
  \fill[orcidlogocol] svg{M256,128c0,70.7-57.3,128-128,128C57.3,256,0,198.7,0,128C0,57.3,57.3,0,128,0C198.7,0,256,57.3,256,128z};
  \fill[white] svg{M86.3,186.2H70.9V79.1h15.4v48.4V186.2z}
  svg{M108.9,79.1h41.6c39.6,0,57,28.3,57,53.6c0,27.5-21.5,53.6-56.8,53.6h-41.8V79.1z M124.3,172.4h24.5c34.9,0,42.9-26.5,42.9-39.7c0-21.5-13.7-39.7-43.7-39.7h-23.7V172.4z}
  svg{M88.7,56.8c0,5.5-4.5,10.1-10.1,10.1c-5.6,0-10.1-4.6-10.1-10.1c0-5.6,4.5-10.1,10.1-10.1C84.2,46.7,88.7,51.3,88.7,56.8z};
 }
}

\newcommand\orcidicon[1]{\href{https://orcid.org/#1}{\mbox{\scalerel*{
    \begin{tikzpicture}[yscale=-1,transform shape]
    \pic{orcidlogo};
    \end{tikzpicture}
   }{|}}}}

\usepackage{graphicx}	
\usepackage{amsmath}	
\usepackage{amssymb}	
\usepackage{multicol}  
\usepackage{bm}		
\usepackage{pdflscape}	
\usepackage{dcolumn}  
\usepackage{array} 
\usepackage{lipsum,float}
\usepackage{subfig, caption}
\usepackage{url}
\usepackage{natbib}
\usepackage{mathtools}
\usepackage[T1]{fontenc}
\usepackage{ae,aecompl}
\usepackage{txfonts}
\usepackage[normalem]{ulem}
\usepackage{booktabs}
\usepackage{epsfig}
\usepackage{hyperref}
\hypersetup{
 colorlinks=true,
 citecolor=black,
 linkcolor=green,
 urlcolor=magenta}

\usepackage[T1]{fontenc}
\usepackage{ae,aecompl}

\usepackage{txfonts}

\title[]{EDGES signal in presence of magnetic-fields}

\author[P. K. Natwariya and J. R. Bhatt]{Pravin Kumar Natwariya \orcidicon{0000-0001-9072-8430}\,$^{1,2}$\thanks{E-mail: \href{mailto:pravin@prl.res.in}{pravin@prl.res.in}} and Jitesh R. Bhatt \orcidicon{0000-0001-7465-8292}\,$^{1}$\thanks{E-mail: \href{mailto:jeet@prl.res.in}{jeet@prl.res.in}}
\\
$^{1}$Physical Research Laboratory, Theoretical Physics Division, Ahmedabad 380 009, India\\
$^{2}$Indian Institute of Technology, Gandhinagar, Ahmedabad 382 424, India}

\date{\today}

\pubyear{}

\begin{document}
\label{firstpage}
\pagerange{\pageref{firstpage}--\pageref{lastpage}}
\maketitle

\begin{abstract}
We study the 21-cm differential brightness temperature in the presence of primordial helical magnetic fields for redshift $z=10-30$. We argue that the $\alpha$-effect that sets in at {\it earlier time } can be helpful in lowering the gas temperature to 3.2 degrees Kelvin at $z=17$. This effect can arise in the early Universe due to some parity violating high energy processes. Using the EDGES (Experiment to Detect the Global Epoch of Reionization Signature) results, we find the upper and lower limits on the primordial magnetic field to be $6\times 10^{-3}~{\rm nG} $ \& $5\times 10^{-4}~{\rm nG}$ respectively. We also discuss the effect of Ly$\alpha$ background on the bounds. Our results do not require any new physics in terms of dark matter.
\end{abstract}

\begin{keywords}
EDGES observation, Magnetic fields, 21-cm signal, Magnetohydrodynamics, first stars, dark matter
\end{keywords}



\begingroup
\let\clearpage\relax
\endgroup
\newpage

\section{Introduction}
Recently, the observations from the Experiment to Detect the Global Epoch of Reionization Signature (EDGES) has created enormous interest in 21-cm cosmology with a hope to provide an insight into the period when the first stars and galaxies were formed (\cite{Bowman:2018yin,Pritchard_2012}). The EDGES collaboration has reported nearly two times more absorption for the 21-cm line than the prediction made by the standard cosmological scenario based on the $\Lambda$CDM framework in the redshift range $15 \lesssim z \lesssim 20$ (\cite{Bowman:2018yin}). Analysis of the results shows that the absorption profile is in a symmetric ``U" shaped form centered at $78\pm 1$~MHz. Minimum of the absorption profile reported being at $-0.5_{-0.5}^{+0.2}$~K in the above-mentioned redshift range. Inability of the standard scenario to explain the observations indicates a possibility of new physics. Any possible explanation may require that either the gas temperature, $T_{\rm gas}$, should be less than $3.2$~K for the standard cosmic microwave background radiation temperature ($T_{\rm CMB}$ ) or $T_{\rm CMB}$ should be grater than 104~K in the absence of any non-standard mechanism for the evolution of the $T_{\rm gas}$ at the centre of the ``U" profile for the best fitting amplitude (\cite{Bowman:2018yin}).

First, it ought to be noted that in the standard cosmological scenario, during the cosmic dawn, $T_{\rm gas}$ and $T_{\rm CMB}$ varies adiabatically with the redshift as $T_{\rm gas}\propto(1+z)^2$ and $T_{\rm CMB}\propto(1+z)$. At redshift $z=17$, temperatures of both the components found to be $T_{\rm gas}\sim 6.8$~K \& $T_{\rm CMB}\sim 48.6$~K, for example see Ref. (\cite{Seager1999}). As explained above, one of the alternatives to explain the EDGES signal is by cooling the gas. In Refs. (\cite{Tashiro:2014tsa, Barkana:2018lgd}), a Coulomb-like interaction between the dark-matter and baryon was considered for transferring energy from gas to dark matter. This approach as argued in Ref. (\cite{Munoz2018}), can violate constraints on local dark-matter density. At the required redshift ionization fraction is the order of, $x_e=n_e/n_H\sim10^{-4}$. Therefore, the dominating part is neutral hydrogen and it possesses only dipole interactions instead of Coulomb-like interaction (\cite{FRASER2018159,Bransden1958}). In addition, the non-standard Coulombic interaction between dark matter and baryons is strongly constrained by observations and laboratory experiments. In the light of these constraints, it is doubtful that one can produce 21-cm absorption signal using the Coulombic interaction (\cite{Barkana:2018nd,Berlin2018,Kovetz2018,Munoz2018a,Slatyer2018}). A new approach was recently adopted in Refs. (\cite{Mirocha2019, Ghara2019}) for the excess cooing of gas by introducing a new parametric model. This model allows the cooling to occur more rapidly at earlier times. However, the origin of the new cooling term remains uncertain. The excessive cooing of the gas can also be obtained by allowing thermal contact between baryons and cold dark matter-axions (\cite{SIKIVIE2019100289}).

Another alternative to explain EDGES results requires extra radiation at the time of cosmic down. This possibility has been investigated by several authors. In Ref. (\cite{FRASER2018159}), the authors consider extra radiation field in the required frequency range by light dark-matter decay into soft photons. In presence of the intergalactic magnetic fields, axion-like particles can be converted into photons under some resonant condition to generate the extra radiation (\cite{Moroi2018}). Similarly, resonant conversion of mirror neutrinos into visible photons can explain the EDGES observations (\cite{AristizabalSierra2018}). In Ref. (\cite{Ewall-Wice2018}), it was suggested that black-holes growing at certain rates can also produce a radio background at the required redshift. However, this type of scenario of the first-stars and black-holes producing enough background radiation was questioned in Refs. (\cite{Sharma2018,Mirocha2019}). 

In this work, we explore a novel possibility of cooling the gas by invoking the so-called alpha-effect. In a conventional plasma, the alpha-effect occurs due to the twisting of magnetic field lines by eddies generated due to the turbulence (\cite{Sur2008,Brandenburg2007}). Here we note that magnetohydrodynamics (MHD) has been studied in the earlier literature from the time of recombination (around redshift $z\sim 1100$) to a very late period of time. In these work, the authors have studied decay of the primordial magnetic field by turbulent decay and the ambipolar diffusion (\cite{Sethi:2004pe,Chluba2015,Minoda:2018gxj}). In turbulence, the twisting of magnetic field lines by eddies in absence of mirror symmetry can enhance the magnetic field. This would give rise to the alpha-effect (\cite{Sur2008,Brandenburg2007}) and the magnetic field enhances at the cost of gas energy. In the present work, we demonstrate  that inclusion of this new effect can change  slope of  gas temperature and thereby it may help in explaining the EDGES signal. In the early Universe such an effect arises due to parity-violating process over a very wide energy scales (\cite{Joyce:1997S,Giovannini:1998S, Pandey:2015kaa, Yamamoto:2016, Boyarsky:2012FR}). This can give rise to helical primordial magnetic fields. Here, we note that alpha-effect may not require any new physics in terms of dark matter. However, the presence of a helical magnetic field is required. Indeed, the primordial magnetic fields (PMFs) generated in the early Universe due to some high energy process may have helical behaviour and violation of parity (\cite{ Joyce:1997S,Giovannini:1998S, Pandey:2015kaa}). These fields can survive in later times (\cite{Boyarsky:2012FR, Pandey:2015kaa}). We believe that this effect can contribute positively to explain the EDGES observations. Additionally, this 21-cm absorption signal can be used as a probe for PMFs strength at the present time in the Universe. In the previous studies, upper bound on the strength of the magnetic fields is constrained for the various cosmological scenarios (for a detailed review see Refs. \cite{Kronberg:1994pp, Neronov:1900zz,Trivedi:2012ssp, Sethi:2004pe, Cheng:1996vn, Grasso:2000wj, Neronov:1900zz, Ade:2015cva, Tashiro:2005ua, Matese:1969cj, Greenstein:1969}). In the context of EDGES signal, constraints on the magnetic fields (MFs) with upper bound of $\lesssim 10^{-10}$~G has been studied by authors of the Ref. (\cite{Minoda:2018gxj}). By invoking baryon dark-matter interaction this upper bound modifies to $\lesssim 10^{-6}$~G (\cite{Bhatt2019pac}). Also, the lower bound on the magnetic field strength found in Refs. (\cite{Ellis:2019MMVA,Fermi_LAT:2018AB,Tavecchio:2010GFB}). In the present work, we argue that the presence of helical magnetic fields and a lack of mirror symmetry implied by the earlier work (\cite{Joyce:1997S,Giovannini:1998S, Pandey:2015kaa, Yamamoto:2016, Boyarsky:2012FR}) can give rise to the alpha-effect around redshift $z\sim 1000$ together with the turbulent decay considered in Refs. (\cite{Minoda:2018gxj, Sethi:2004pe}). 

To compute the 21-cm differential brightness temperature, $T_{21}$, we use \href{http://homepage.sns.it/mesinger/DexM___21cmFAST.html}{21cmFAST} code. We modify this code by adding `decay' rates related with turbulence and ambipolar effects associated with the magnetic field together with the alpha effect. Following definition of $T_{21}$ given in Refs. (\cite{Furlanetto2006a,Mesinger:2007S,Mesinger:2011FS}),
we write, 
\begin{alignat}{2}
 T_{21}=27x_{\rm HI}\frac{1}{1+\partial_r v_r/H}&(1+\delta_{\rm nl})\left(\frac{\Omega_{\rm M }h^2}{0.15}\right)^{-1/2}\left(\frac{\Omega_{\rm b}h^2}{0.023}\right)\nonumber\\
 \times & \left(\frac{1+z}{10}\right)^{1/2}\left(1-\frac{T_{\rm CMB}}{T_s}\right)~{\rm mK}\,,\label{eq-1}
\end{alignat}
where, $x_{\rm HI}$ is the neutral hydrogen fraction, $\partial_r v_r$ is the comoving derivative of LOS component of the comoving velocity, $H\equiv H(z)$ is the Hubble expansion rate and $\delta_{\rm nl}\equiv \delta_{\rm nl}(\bm x,z) $ is the density contrast. We take the following values for the cosmological parameters: $\Omega_{\rm M }=0.31$, $\Omega_{\rm b}=0.048$, $h=0.68$, $\sigma_8=0.82$, $n_s=0.97$ and $T_{\rm CMB}|_{z=0}=T_0=2.726$~K (\cite{Planck:2018,Fixsen_2009}). The spin temperature $T_s$ is defined via hydrogen number densities in $1\rm S$ triplet ($n_1$) and singlet ($n_0$) hyperfine levels: ${n_1/n_0}={g_1/g_0}\times \exp(-2\pi\nu_{10}/T_s)\,,$ here, $g_1$ and $g_0$ are spin degeneracy in triplet and singlet states respectively and $\nu_{10}$ is corresponding frequency for hyperfine transition. We write $T_s $ (\cite{Furlanetto2006a,Pritchard_2012}),
\begin{alignat}{2}
 T_s^{-1}=\frac{T_{\rm CMB}^{-1} + x_\alpha T_\alpha^{-1}+x_cT_{\rm gas}^{-1} }{1+x_\alpha+x_c}\,\label{eq-2}.
\end{alignat}
Here, $T_\alpha\approx T_{\rm gas}$ is the colour temperature (\cite{1952AJ.....57R..31W,Field}). $x_\alpha$ and $x_c$ are Wouthuysen-Field (WF) and collisional coupling coefficients respectively (\cite{1952AJ.....57R..31W,Field,Hirata2006,Mesinger:2011FS}). We consider that first stars were formed at redshift $z\sim30$. Later, their Ly$\alpha$ background can cause the hyperfine transition and X-ray produced by these sources start to heat the gas (\cite{Furlanetto2006a,Ghara2019,Mirocha2019,Mesinger:2011FS,Mesinger2013a,Fialkov:2016A,Park:2019}). For this work, we take the fiducial model as defined in the Ref. (\cite{Mesinger:2011FS}). Following the above References, we switch on the effect of Lyman $\alpha$ background and structure formation on $T_{\rm gas}$ after $z=30$. It is important to note here that in Ref. (\cite{PhysRevD.98.103513}), the authors have claimed that $T_{\rm gas}$ values can be even higher, without X-ray heating, if one incorporates indirect energy transfer from radio photons to the random motions of the gas. 

In the presence of magnetic fields thermal evolution of the gas can modify. We follow the Refs. (\cite{Shu:1992fh, Sethi:2008eq, Schleicher:2008aa,Sethi:2004pe, Chluba2015}), and write the temperature evolution of the gas in presence of PMFs as,
\begin{alignat}{2}
 \frac{dT_{\rm gas}}{dz}=2\,\frac{T_{\rm gas}}{1+z}+\frac{\Gamma_c}{(1+z)\,H}&(T_{\rm gas}-T_{\rm CMB})\nonumber\\
 &-\frac{2\,(\Gamma_{\rm turb}+\Gamma_{\rm ambi}+\Gamma_{\rm alph})}{3\,N_{\rm tot}(1+z)\,H}\,,\label{eq1}
\end{alignat}
where, $N_{\rm tot}$ is the total number density of the gas i.e. $N_{\rm H}\,(1+f_{\rm He}+X_e)$, $N_{\rm H}$ is the neutral hydrogen number density, $f_{\rm He}\approx \frac{Y_p}{4\,(1-Y_p)}$, Helium mass fraction $Y_p=0.24$, $X_e=N_e/N_{\rm H}$ is the free streaming electron fraction in the gas, $\Gamma_c$ is the Compton scattering rate (\cite{Schleicher:2008aa,Chluba2015}). To get free electron fraction, $X_e$, we follow (\cite{Seager1999,Seager}) and correction suggested by (\cite{Chluba2010a,Chluba2010,Hart2018}). In equation \eqref{eq1}, $\Gamma_{\rm turb}$, $\Gamma_{\rm ambi}$ and $\Gamma_{\rm alph}$ are heating or cooling rate per unit volume due to the turbulence, ambipolar and alpha effect respectively. $\Gamma_{\rm ambi}$ and $\Gamma_{\rm turb}$ are (\cite{Sethi:2004pe, Chluba2015}),
\begin{alignat}{2}
 &\Gamma_{\rm ambi}\approx\frac{(1-X_e)}{\gamma\, X_e\, (M_HN_b)^2}\ \frac{E_B^2\,f_L(n_B+3)}{L_d^2}\,,\\
 &\Gamma_{\rm turb}=\frac{1.5\ m\ \left[\ln(1+t_i/t_d)\right]^m}{\left[\ln(1+t_i/t_d)+1.5\ln\{(1+z_i)/(1+z)\}\right]^{m+1}}H\,E_B\,,
\end{alignat}
\begin{figure*}
	\centering
	\subfloat[] {\includegraphics[width=3.5in,height=2.2in]{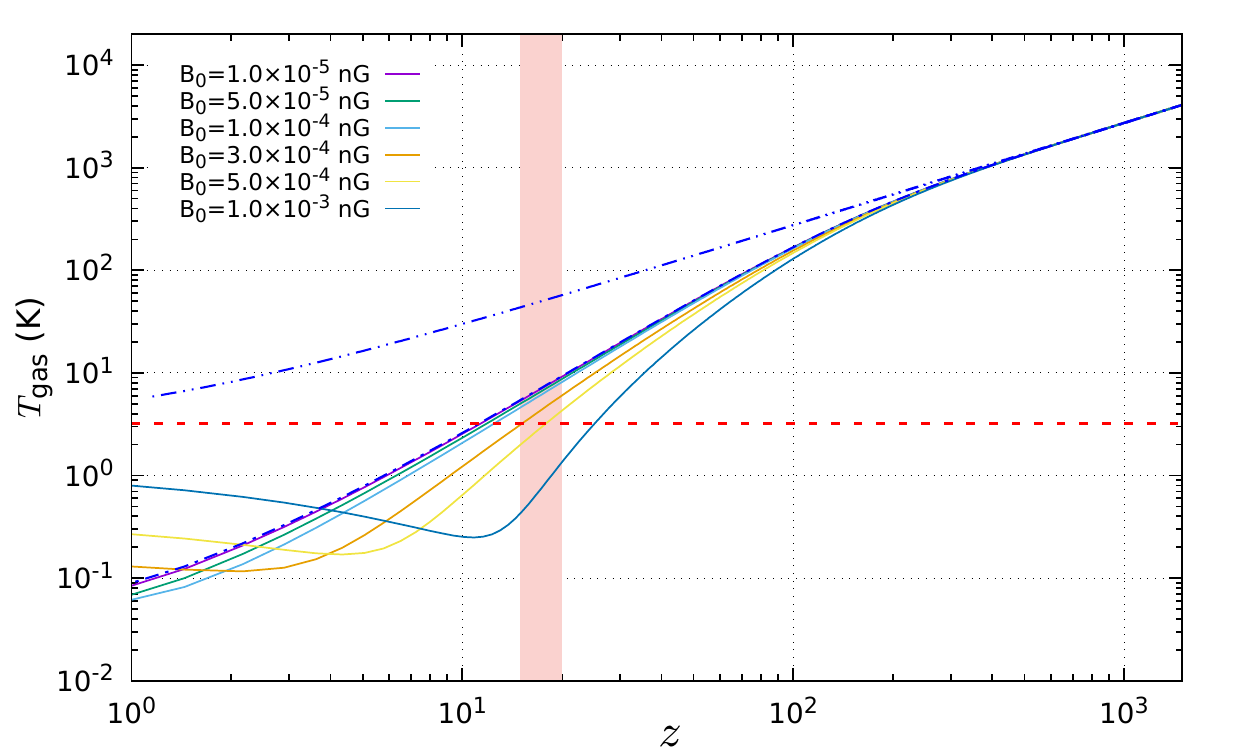}\label{plot:1a}}
	\subfloat[] {\includegraphics[width=3.5in,height=2.2in]{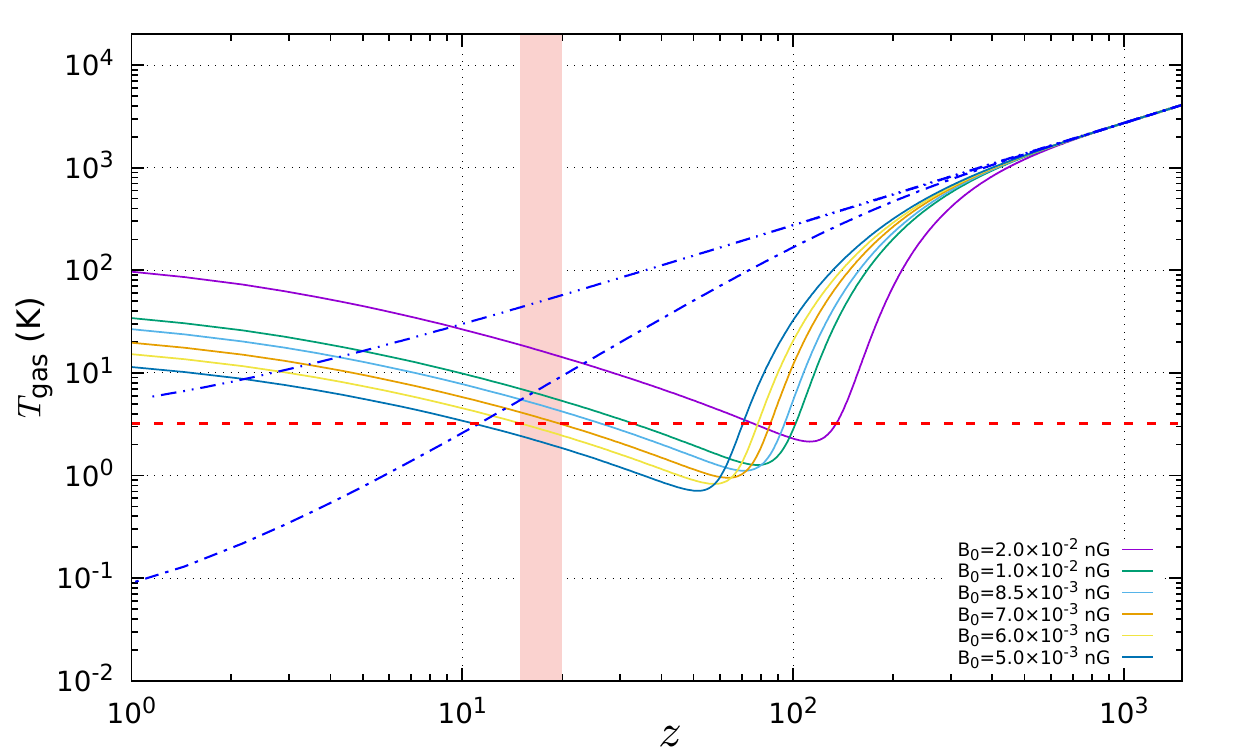}\label{plot:1b}} 
	\caption{The gas temperature evolution with redshift for different magnetic field strengths -- solid lines. The blue dot-dashed line indicates the $T_{\rm gas}$ evolution for the standard cosmological scenario and double-dot dashed line shows $T_{\rm CMB}$ evolution. The shaded region is corresponds to 21-cm absorption signal, $15\leq z \leq 20$, reported by EDGES observation. The red dashed horizontal line is corresponds to the $T_{\rm gas}=3.2$~K. }
	\label{plot:1detail}
\end{figure*} 
here, the coupling coefficient $\gamma=1.9\times 10^{14}\,(T_{\rm gas}/{\rm K})^{0.375}{\rm cm}^3$ $/{\rm g}/{\rm s}$, $M_{\rm H}$ is mass of Hydrogen atom, $N_b$ is baryon number density, $f_L(x)=0.8313\,(1-1.020\times10^{-2}x)\,x^{1.105}$, $t_i/t_d\approx14.8\,(1+z)\,({\rm nG}/B_0)\,(L_d/{\rm Mpc})$, $m=2(n_B+3)/(n_B+5)$, $z_i=1088$ is the initial redshift when heating starts \& $L_d$ is the coherence length scale of the magnetic field.  It  is constrained by Alfv\'{e}n wave damping length scale, $L_d=1/[k_{\rm d}\,(1+z)]$. Magnetic fields at length-scales smaller than $L_d$ are strongly damped by the radiative-viscosity  (\cite{Sethi:2004pe,Jedamzik1998,Chluba2015,Kunze_2014}),
\begin{alignat}{2}
k_{\rm d}\simeq 286.91\,\left(\frac{\rm nG}{B_0}\right)\, {\rm Mpc}^{-1}\, \label{eq6}
\end{alignat}
where, $B_0$ is the present day magnetic field strength.
\begin{figure}
	\includegraphics[width=3.4in,height=2.1in]{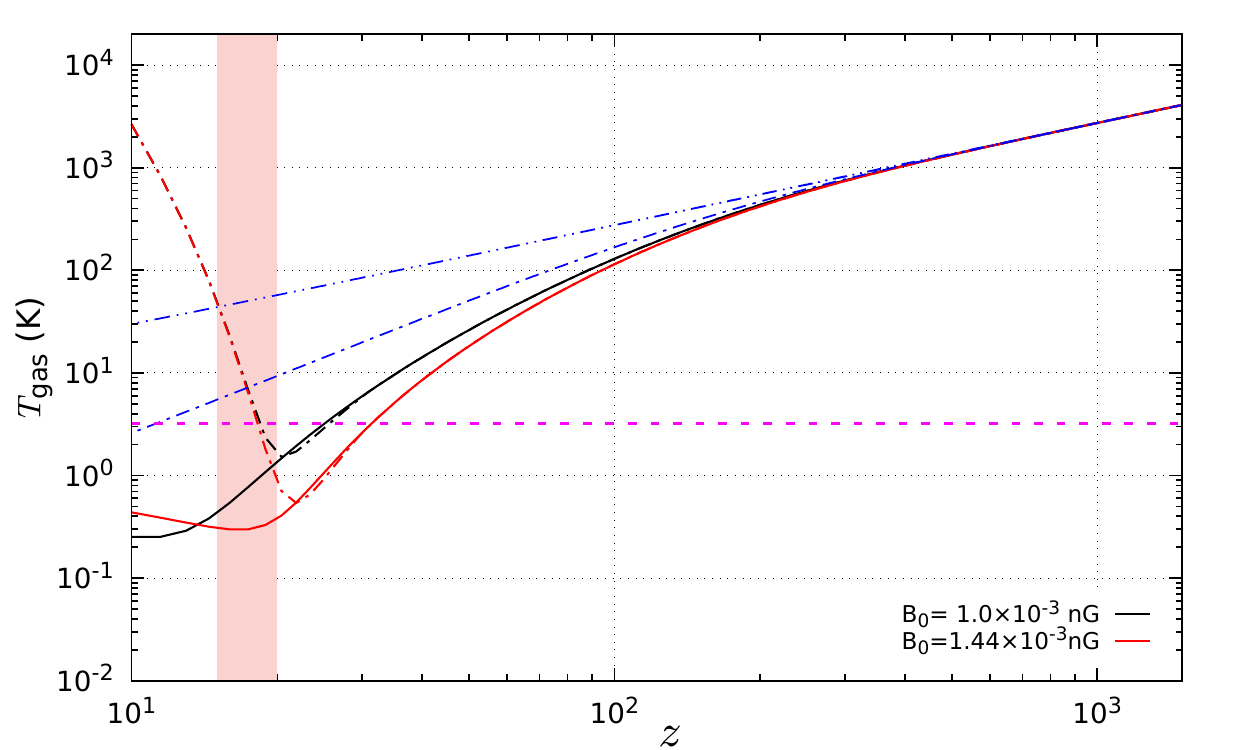}\label{x_ray}
	\caption{Black (red) solid line represent $T_{\rm gas}$ evolution with $z$ in presence of magnetic field. Dot-dashed lines indicate the X-ray heating for different magnetic field strengths. The blue coloured lines represent the standard cosmological scenario. The magenta dashed line is corresponds to the $T_{\rm gas}=3.2$~K. }
	\label{plot:3detail}
\end{figure}
%
%

The process of inverse cascade in presence of the $\alpha$-effect has been studied in cosmology literature where magnetic energy flows from smaller length scales to the larger length scale (\cite{PhysRevE.64.056405,PhysRevD.96.023504,OLESEN1997321}). It is found that  the typical comoving length scales at which the coherent  primordial magnetic field can exist is around few Mpc . Strength of PMFs is constrained by Planck collaboration on the length scale of 1~Mpc (\cite{2014A&A...571A..16P}). We consider power-spectrum of the PMFs as a power law  $P_B(k)=Ak^{n_B}$ for $k<k_d$ and $P_B(k)=0$ for $k\geq k_d$. Amplitude of the PMFs smoothed over length scale $\lambda$, $B_\lambda^2=\int_{0}^{\infty}(dk/2\pi)^3\,P_{B,0}(k)\exp(-k^2\lambda^2)=\big(\sqrt2/(k_d\lambda)\big)^{n_B+3}B_0^2$ (\cite{PhysRevD.69.063006,Chluba2015,2014A&A...571A..16P}). For a nearly scale invariant magnetic spectral index, $n_B=-2.9$  and $\lambda=1$~Mpc, using the above relation for $B_\lambda^2$,  one can estimate the amplitude of PMFs at 1~Mpc ($B_{1\,\rm Mpc}$) (\cite{Chluba2015}). The magnetic field energy density, $E_B=B^2/(8\pi)$, in presence of the alpha-effect,
\begin{alignat}{2}
 \frac{dE_B}{dz}=4\,\frac{E_B}{1+z}+\frac{1}{(1+z)\,H}\,&\Big[\,\Gamma_{\rm turb}+\Gamma_{\rm ambi}\nonumber\\
 &-\frac{\alpha}{4\,\pi}\,\big|\bm{B}\,.\,(\,\bm{\nabla}\times \bm{B}\,)\,\big|\,\Big]\,.\label{eq4}
\end{alignat}
Here, $ \bm B\equiv \langle{\bm B}\rangle$. Following Refs. (\cite{Sur2008,Brandenburg2007,Brandenburg2005a}), if the magnetic Reynolds number is large enough, $\alpha=(1/3)\,u_{\rm rms}$. Here we use Equipartition theorem-- $u_{\rm rms}^2=3T_{\rm gas}/M_{\rm H}$. Following Ref. (\cite{Schleicher:2008aa,PhysRevD.69.063006}), we approximate last term in equation \eqref{eq4} as
\begin{alignat}{2}
 \big|\bm{B}\,.\,(\,\bm{\nabla}\times \bm{B}\,)\,\big| \approx \frac{B^2}{L_d}\,.\label{eq5} 
\end{alignat}
 Thus, in equation \eqref{eq4},
\begin{equation}
 \Gamma_{\rm alph}\simeq-2\ \left(\frac{T_{\rm gas}}{3\,M_{\rm H}}\right)^{1/2}\, \frac{E_B}{L_d}\,.\label{eq7}
\end{equation}
Here we note that in addition to the usual expansion term in Eq. \eqref{eq4}, the terms with coefficients $\Gamma_{\rm turb}$ and $\Gamma_{\rm ambi}$  also contribute towards the decay of magnetic fields. But, the term with factor $\alpha$ has sign opposite to the decay terms and this will help the magnetic field to survive for a longer duration. However, the decay terms will eventually dominate over the $\alpha$-effect.
 


\section{result and discussion} 
Ignoring logarithmic dependency of turbulent decay, it evolves as $\Gamma_{\rm turb}\propto (1+z)^{5.2}$, ambipolar diffusion $\Gamma_{\rm ambi}\propto (1+z)^{3.63}(1-X_e)/X_e$ at early time since $T_{\rm gas}\propto (1+z)$ and after $z\lesssim 100$ it evolves as $\propto(1+z)^{3.25}/X_e$ because of $T_{\rm gas}\propto (1+z)^2$ and $X_e\ll 1$ at late time. Magnetic energy rate due to the alpha-effect, $\Gamma_{\rm alph}$, is $\propto (1+z)^{5.5}$ for $z\gtrsim 100$ otherwise it's $\propto(1+z)^6$. Therefore, we expect cooling due to the alpha-effect is more effective than heating due to the turbulent decay. After that, at late time ($z<100$) the ambipolar diffusion is more effective (also depends on PMFs strength).  Thus, the gas temperature  will fall quickly in comparison with the standard scenario--heating/cooling due to magnetic fields is not included. As shown in Refs. (\cite{Minoda:2018gxj, Chluba2015}) in presence of a helical magnetic field $\Gamma_{\rm turb}$ dominates over $\Gamma_{\rm ambi}$ for $z>100$. Presence of the alpha effect can also be felt very strongly for this range of the redshift. One can write $\frac{\Gamma_{\rm alph}}{\Gamma_{\rm ambi}} \sim 1.48\left(\frac{T_{\rm gas}}{\rm Kelvin}\right)^{0.875}\frac{x_e}{1-x_e}\,(1+z)\,\left( \frac{\rm nG}{B_0} \right)$. The growth in the magnetic energy density due to $\alpha$-effect in the present scenario can be estimated as follows: From equation \eqref{eq1} and \eqref{eq4}, 
 \begin{equation}
\frac{dT_{\rm gas}}{dz}=\frac{dT_{\rm gas}}{dz}\bigg|_{\rm std}-\frac{2}{3N_{\rm tot}}\left[\frac{dE_B}{dz}-\frac{dE_B}{dz}\bigg|_{\rm std}\right]\,,\label{eq10}
 \end{equation}
 where, $\frac{dT_{\rm gas}}{dz}\big|_{\rm std}$ represent the gas temperature evolution with redshift in absence of the magnetic fields. While,  $\frac{dE_B}{dz}\big|_{\rm std}=4\,\frac{E_B}{1+z}\,,$ represent the magnetic field energy density evolution with redshift without any $\alpha$-effect, turbulent decay and ambipolar diffusion. For example, in figure \eqref{plot:1detail}, maximum transformation of thermal energy to magnetic energy happens at $z\approx58$ for $B_0=6\times10^{-3}$~nG. Now using equation \eqref{eq10} and results of plot \eqref{plot:1detail} one can find $E_B$ and which gives $B\,\big|_{z\approx 58} \approx 4 \times 10^{-7}$~G. The upper constraint from Planck results on the present day value of magnetic field is around $4.1\times 10^{-9}$~G, from which one can estimate magnetic field at $z=58$ around $1.4\times 10^{-5}$~G. Thus, the magnetic field generated by the $\alpha$-effect is consistent with Planck bound (\cite{2014A&A...571A..16P}). 

To study the magnetic heating (cooling) of the gas we use the code \href{http://www.jb.man.ac.uk/~jchluba/Science/CosmoRec/Recfast++.html}{recfast++} (\cite{Chluba2015}). In figure \eqref{plot:1a}, plots of gas temperatures for different values of $B_0$ are shown as function of $z$. The dot-dashed line represent the standard recombination history. The figure shows that as values of $B_0$ approaches $10^{-5}$~nG, $T_{\rm gas}$ recovers the standard thermal evolution. By increasing magnetic field from $10^{-5}$~nG, $T_{\rm gas}$ decreases. For $B_{0}\approx 10^{-3}$~nG, one gets $T_{\rm gas}<$~3.2 Kelvin for $z=17$. Further we note that by increasing $B_0$ the minimum of gas temperature shifts towards higher values of the redshift. Figure \eqref{plot:1b} shows that by increasing of $B_0$ from $5\times 10^{-3}$~nG, $T_{\rm gas}$ rises. However, gas temperature around $z=17$ exceeds 3.2 Kelvin for $B_0 > 6\times 10^{-3}~{\rm nG}$. Therefore, desired value of magnetic field should be in the range of $5\times10^{-4}~{\rm nG}\lesssim B_0 \lesssim 6\times 10^{-3}~{\rm nG}$. These upper and lower bounds on $B_0$ are also consistent with constraints found in Refs. (\cite{Kronberg:1994pp, Neronov:1900zz,Trivedi:2012ssp, Sethi:2004pe, Cheng:1996vn, Grasso:2000wj, Neronov:1900zz,Ade:2015cva,Tashiro:2005ua, Matese:1969cj, Greenstein:1969, Minoda:2018gxj, Bhatt2019pac,Ellis:2019MMVA,Fermi_LAT:2018AB,Tavecchio:2010GFB}).

Further, in figure (\ref{plot:3detail}), we have included the X-ray heating due to first stars after the redshift $z=30$ together with the adiabatic heating/cooling as a result of structure formation. The blue dot-dashed line indicates the $T_{\rm gas}$ evolution for the standard cosmological scenario and double-dot dashed line shows $T_{\rm CMB}$ evolution. The black and red solid line plots represent the case when only magnetic heating/cooling terms are included. The black and red dot-dashed line shows the cases when all these effects are present. In this case, the gas temperature rises quickly in comparison with the only magnetic heating/cooling cases. Therefore, in the presence of X-ray heating, our previously mentioned upper and lower bounds on magnetic field strength can modify.

\begin{figure}
	\includegraphics[width=3.5in,height=2.2in]{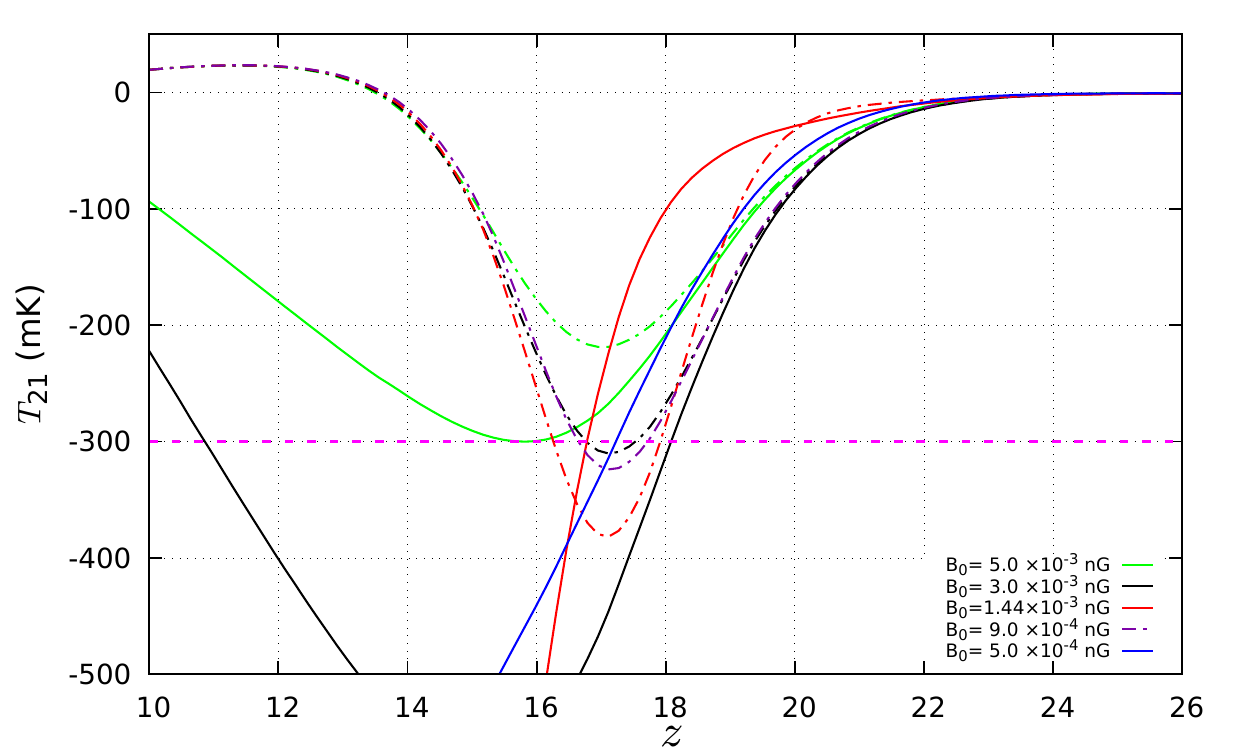}\label{T21}
	\caption{21-cm global signal in presence (dot-dashed) and absence (solid lines) of X-ray heating of gas for different magnetic field strengths. The magenta dashed line is corresponds to the EDGES upper bound on $T_{21}: -300$~mK.}
	\label{plot:2detail}
\end{figure}

In figure (\ref{plot:2detail}), we plot $T_{21}$ as a function of redshift for different magnetic field strengths. We have considered two particular cases involving (dot-dashed lines) and without (solid lines) X-ray heating. For the X-ray heating we consider the fiducial model (\cite{Mesinger:2011FS}). In all cases, we incorporate adiabatic heating/cooling from structure formations (\cite{Mesinger:2011FS}). Spin temperature coupling has two main contribution, one from X-ray excitation of neutral hydrogen and other from photons emitted between Lyman to Ly$\alpha$ limit from the first stars. For dot-dashed line we include both coupling and for solid lines, we take only second coupling. The figure shows that without including X-ray heating, one can obtain $-1000{~\rm mK}\leq T_{21} \leq -300$~mK. 
For the dot-dashed lines, the minimum of $T_{21}$ profile first decreases while increasing $B_0$ values from $\sim 9\times 10^{-4}$~nG and after a certain value of $B_0$ , minimum of $T_{21}$ starts increasing: For $B_0= 3\times 10^{-3}$~nG and $9\times 10^{-4}$~nG we get $T_{21}=-310$~mK and $-323$~mK at $z=17$ respectively. This gives allowed range for $B_0$ to be in the range (using EDGES upper bound on $T_{21}$) $9\times10^{-4}~{\rm nG}\lesssim B_0 \lesssim 3\times 10^{-3}~{\rm nG}$ after inclusion of X-ray heating.


\section{conclusions}
In conclusion, we have studied 21-cm differential brightness temperature in the presence of helical primordial magnetic fields. We have shown that the presence of the alpha effect can reduce gas temperature to 3.2 Kelvin, at the center of ``U" shaped profile, when present-day strength of the magnetic field is in the range $5\times10^{-4}~{\rm nG}\lesssim B_0 \lesssim 6\times 10^{-3}~{\rm nG}$ without X-ray heating. For the case when X-ray heating is included we get $9\times10^{-4}~{\rm nG}\lesssim B_0 \lesssim 3\times 10^{-3}~{\rm nG}$ for $T_{21}\lesssim-300$~mK. Here we note that our analysis does not require any new physics in terms of dark matter. 

\section*{Acknowledgements}
\addcontentsline{toc}{section}{Acknowledgements}
We would like to thank the anonymous referee whose comments has helped us in improving presentation of our results. All the computations were performed on the Vikram-100 HPC cluster at PRL, Ahmedabad. 




\bibliographystyle{mnras}

\begin{thebibliography}{}
	\makeatletter
	\relax
	\def\mn@urlcharsother{\let\do\@makeother \do\$\do\&\do\#\do\^\do\_\do\%\do\~}
	\def\mn@doi{\begingroup\mn@urlcharsother \@ifnextchar [ {\mn@doi@}
		{\mn@doi@[]}}
	\def\mn@doi@[#1]#2{\def\@tempa{#1}\ifx\@tempa\@empty \href
		{http://dx.doi.org/#2} {doi:#2}\else \href {http://dx.doi.org/#2} {#1}\fi
		\endgroup}
	\def\mn@eprint#1#2{\mn@eprint@#1:#2::\@nil}
	\def\mn@eprint@arXiv#1{\href {http://arxiv.org/abs/#1} {{\tt arXiv:#1}}}
	\def\mn@eprint@dblp#1{\href {http://dblp.uni-trier.de/rec/bibtex/#1.xml}
		{dblp:#1}}
	\def\mn@eprint@#1:#2:#3:#4\@nil{\def\@tempa {#1}\def\@tempb {#2}\def\@tempc
		{#3}\ifx \@tempc \@empty \let \@tempc \@tempb \let \@tempb \@tempa \fi \ifx
		\@tempb \@empty \def\@tempb {arXiv}\fi \@ifundefined
		{mn@eprint@\@tempb}{\@tempb:\@tempc}{\expandafter \expandafter \csname
			mn@eprint@\@tempb\endcsname \expandafter{\@tempc}}}
	
	\bibitem[\protect\citeauthoryear{Ade et~al.}{Ade et~al.}{2016}]{Ade:2015cva}
	Ade P. A.~R.,  et~al., 2016, \mn@doi [A\&A] {10.1051/0004-6361/201525821}, 594,
	A19
	
	\bibitem[\protect\citeauthoryear{{Aristizabal Sierra} \& Fong}{{Aristizabal
			Sierra} \& Fong}{2018}]{AristizabalSierra2018}
	{Aristizabal Sierra} D.,  Fong C.~S.,  2018, \mn@doi [PLB]
	{10.1016/J.PHYSLETB.2018.07.047}, 784, 130
	
	\bibitem[\protect\citeauthoryear{Barkana}{Barkana}{2018}]{Barkana:2018lgd}
	Barkana R.,  2018, \mn@doi [Nature] {10.1038/nature25791}, 555, 71
	
	\bibitem[\protect\citeauthoryear{Barkana, Outmezguine, Redigolo  \&
		Volansky}{Barkana et~al.}{2018}]{Barkana:2018nd}
	Barkana R.,  Outmezguine N.~J.,  Redigolo D.,   Volansky T.,  2018, \mn@doi
	[PRD] {10.1103/PhysRevD.98.103005}, 98, 103005
	
	\bibitem[\protect\citeauthoryear{Berlin, Hooper, Krnjaic  \& McDermott}{Berlin
		et~al.}{2018}]{Berlin2018}
	Berlin A.,  Hooper D.,  Krnjaic G.,   McDermott S.~D.,  2018, \mn@doi [PRL]
	{10.1103/PhysRevLett.121.011102}, 121, 011102
	
	\bibitem[\protect\citeauthoryear{Bhatt \& Pandey}{Bhatt \&
		Pandey}{2016}]{Pandey:2015kaa}
	Bhatt J.~R.,  Pandey A.~K.,  2016, \mn@doi [PRD] {10.1103/PhysRevD.94.043536},
	94, 043536
	
	\bibitem[\protect\citeauthoryear{Bhatt, Natwariya, Nayak  \& Pandey}{Bhatt
		et~al.}{2020}]{Bhatt2019pac}
	Bhatt J.~R.,  Natwariya P.~K.,  Nayak A.~C.,   Pandey A.~K.,  2020, \mn@doi
	[Eur. Phys. J. C] {10.1140/epjc/s10052-020-7886-x}, 80, 334
	
	\bibitem[\protect\citeauthoryear{Bowman, Rogers, Monsalve, Mozdzen  \&
		Mahesh}{Bowman et~al.}{2018}]{Bowman:2018yin}
	Bowman J.~D.,  Rogers A. E.~E.,  Monsalve R.~A.,  Mozdzen T.~J.,   Mahesh N.,
	2018, \mn@doi [Nature] {10.1038/nature25792}, 555, 67
	
	\bibitem[\protect\citeauthoryear{Boyarsky, Fr\"ohlich  \& Ruchayskiy}{Boyarsky
		et~al.}{2012}]{Boyarsky:2012FR}
	Boyarsky A.,  Fr\"ohlich J.,   Ruchayskiy O.,  2012, \mn@doi [PRL]
	{10.1103/PhysRevLett.108.031301}, 108, 031301
	
	\bibitem[\protect\citeauthoryear{Brandenburg \& Subramanian}{Brandenburg \&
		Subramanian}{2005}]{Brandenburg2005a}
	Brandenburg A.,  Subramanian K.,  2005, \mn@doi [A\&A]
	{10.1051/0004-6361:20053221}, 439, 835
	
	\bibitem[\protect\citeauthoryear{Brandenburg \& Subramanian}{Brandenburg \&
		Subramanian}{2007}]{Brandenburg2007}
	Brandenburg A.,  Subramanian K.,  2007, \mn@doi [Ast. Nac.]
	{10.1002/asna.200710772}, 328, 507
	
	\bibitem[\protect\citeauthoryear{Bransden, Dalgarno, John  \& Seaton}{Bransden
		et~al.}{1958}]{Bransden1958}
	Bransden B.~H.,  Dalgarno A.,  John T.~L.,   Seaton M.~J.,  1958, \mn@doi
	[Proc. Phys. Soc.] {10.1088/0370-1328/71/6/301}, 71, 877
	
	\bibitem[\protect\citeauthoryear{Caprini, Durrer  \& Kahniashvili}{Caprini
		et~al.}{2004}]{PhysRevD.69.063006}
	Caprini C.,  Durrer R.,   Kahniashvili T.,  2004, \mn@doi [Phys. Rev. D]
	{10.1103/PhysRevD.69.063006}, 69, 063006
	
	\bibitem[\protect\citeauthoryear{Cheng, Olinto, Schramm  \& Truran}{Cheng
		et~al.}{1996}]{Cheng:1996vn}
	Cheng B.,  Olinto A.~V.,  Schramm D.~N.,   Truran J.~W.,  1996, \mn@doi [PRD]
	{10.1103/PhysRevD.54.4714}, 54, 4714
	
	\bibitem[\protect\citeauthoryear{Chluba \& Thomas}{Chluba \&
		Thomas}{2011}]{Chluba2010}
	Chluba J.,  Thomas R.~M.,  2011, \mn@doi [MNRAS]
	{10.1111/j.1365-2966.2010.17940.x}, 412, 748
	
	\bibitem[\protect\citeauthoryear{Chluba, Vasil  \& Dursi}{Chluba
		et~al.}{2010}]{Chluba2010a}
	Chluba J.,  Vasil G.~M.,   Dursi L.~J.,  2010, \mn@doi [MNRAS]
	{10.1111/j.1365-2966.2010.16940.x}, 407, 599
	
	\bibitem[\protect\citeauthoryear{Chluba, Paoletti, Finelli  \&
		Rubi{\~{n}}o-Mart{\'{i}}n}{Chluba et~al.}{2015}]{Chluba2015}
	Chluba J.,  Paoletti D.,  Finelli F.,   Rubi{\~{n}}o-Mart{\'{i}}n J.~A.,  2015,
	\mn@doi [MNRAS] {10.1093/mnras/stv1096}, 451, 2244
	
	\bibitem[\protect\citeauthoryear{Christensson, Hindmarsh  \&
		Brandenburg}{Christensson et~al.}{2001}]{PhysRevE.64.056405}
	Christensson M.,  Hindmarsh M.,   Brandenburg A.,  2001, \mn@doi [Phys. Rev. E]
	{10.1103/PhysRevE.64.056405}, 64, 056405
	
	\bibitem[\protect\citeauthoryear{Ellis, Fairbairn, Lewicki, Vaskonen  \&
		Wickens}{Ellis et~al.}{2019}]{Ellis:2019MMVA}
	Ellis J.,  Fairbairn M.,  Lewicki M.,  Vaskonen V.,   Wickens A.,  2019,
	\mn@doi [JCAP] {10.1088/1475-7516/2019/09/019}, 2019, 019
	
	\bibitem[\protect\citeauthoryear{Ewall-Wice, Chang, Lazio, Dor{\'{e}}, Seiffert
		\& Monsalve}{Ewall-Wice et~al.}{2018}]{Ewall-Wice2018}
	Ewall-Wice A.,  Chang T.-C.,  Lazio J.,  Dor{\'{e}} O.,  Seiffert M.,
	Monsalve R.~A.,  2018, \mn@doi [ApJ] {10.3847/1538-4357/aae51d}, 868, 63
	
	\bibitem[\protect\citeauthoryear{Fialkov, Cohen, Barkana  \& Silk}{Fialkov
		et~al.}{2016}]{Fialkov:2016A}
	Fialkov A.,  Cohen A.,  Barkana R.,   Silk J.,  2016, \mn@doi [MNRAS]
	{10.1093/mnras/stw2540}, 464, 3498
	
	\bibitem[\protect\citeauthoryear{{Field}}{{Field}}{1958}]{Field}
	{Field} G.~B.,  1958, \mn@doi [Proceedings of the IRE]
	{10.1109/JRPROC.1958.286741}, 46, 240
	
	\bibitem[\protect\citeauthoryear{Fixsen}{Fixsen}{2009}]{Fixsen_2009}
	Fixsen D.~J.,  2009, \mn@doi [ApJ] {10.1088/0004-637x/707/2/916}, 707, 916
	
	\bibitem[\protect\citeauthoryear{Fraser et~al.}{Fraser
		et~al.}{2018}]{FRASER2018159}
	Fraser S.,  et~al., 2018, \mn@doi [PLB] {10.1016/j.physletb.2018.08.035}, 785,
	159
	
	\bibitem[\protect\citeauthoryear{Furlanetto \& Pritchard}{Furlanetto \&
		Pritchard}{2006}]{Furlanetto2006a}
	Furlanetto S.~R.,  Pritchard J.~R.,  2006, \mn@doi [MNRAS]
	{10.1111/j.1365-2966.2006.10899.x}, 372, 1093
	
	\bibitem[\protect\citeauthoryear{Ghara \& Mellema}{Ghara \&
		Mellema}{2019}]{Ghara2019}
	Ghara R.,  Mellema G.,  2019, \mn@doi [MNRAS] {10.1093/mnras/stz3513}, 492, 634
	
	\bibitem[\protect\citeauthoryear{Giovannini \& Shaposhnikov}{Giovannini \&
		Shaposhnikov}{1998}]{Giovannini:1998S}
	Giovannini M.,  Shaposhnikov M.~E.,  1998, \mn@doi [PRD]
	{10.1103/PhysRevD.57.2186}, 57, 2186
	
	\bibitem[\protect\citeauthoryear{Grasso \& Rubinstein}{Grasso \&
		Rubinstein}{2001}]{Grasso:2000wj}
	Grasso D.,  Rubinstein H.~R.,  2001, \mn@doi [Phys. Rept.]
	{10.1016/S0370-1573(00)00110-1}, 348, 163
	
	\bibitem[\protect\citeauthoryear{Greenstein}{Greenstein}{1969}]{Greenstein:1969}
	Greenstein G.,  1969, \mn@doi [Nature] {10.1038/223938b0}, 223, 938
	
	\bibitem[\protect\citeauthoryear{Hart \& Chluba}{Hart \&
		Chluba}{2018}]{Hart2018}
	Hart L.,  Chluba J.,  2018, \mn@doi [MNRAS] {10.1093/mnras/stx2783}, 474, 1850
	
	\bibitem[\protect\citeauthoryear{Hirata}{Hirata}{2006}]{Hirata2006}
	Hirata C.~M.,  2006, \mn@doi [MNRAS] {10.1111/j.1365-2966.2005.09949.x}, 367,
	259
	
	\bibitem[\protect\citeauthoryear{Jedamzik, Katalini\ifmmode~\acute{c}\else
		\'{c}\fi{}  \& Olinto}{Jedamzik et~al.}{1998}]{Jedamzik1998}
	Jedamzik K.,  Katalini\ifmmode~\acute{c}\else \'{c}\fi{} V. c.~v.,   Olinto
	A.~V.,  1998, \mn@doi [Phys. Rev. D] {10.1103/PhysRevD.57.3264}, 57, 3264
	
	\bibitem[\protect\citeauthoryear{Joyce \& Shaposhnikov}{Joyce \&
		Shaposhnikov}{1997}]{Joyce:1997S}
	Joyce M.,  Shaposhnikov M.,  1997, \mn@doi [PRL] {10.1103/PhysRevLett.79.1193},
	79, 1193
	
	\bibitem[\protect\citeauthoryear{Kovetz, Poulin, Gluscevic, Boddy, Barkana  \&
		Kamionkowski}{Kovetz et~al.}{2018}]{Kovetz2018}
	Kovetz E.~D.,  Poulin V.,  Gluscevic V.,  Boddy K.~K.,  Barkana R.,
	Kamionkowski M.,  2018, \mn@doi [PRD] {10.1103/PhysRevD.98.103529}, 98,
	103529
	
	\bibitem[\protect\citeauthoryear{Kronberg}{Kronberg}{1994}]{Kronberg:1994pp}
	Kronberg P.,  1994, \mn@doi [Rep. Prog. Phys] {10.1088/0034-4885/57/4/001},
	\href {http://adsabs.harvard.edu/abs/1994RPPh...57..325K} {57, 325}
	
	\bibitem[\protect\citeauthoryear{Kunze \& Komatsu}{Kunze \&
		Komatsu}{2014}]{Kunze_2014}
	Kunze K.~E.,  Komatsu E.,  2014, \mn@doi [JCAP]
	{10.1088/1475-7516/2014/01/009}, 2014, 009
	
	\bibitem[\protect\citeauthoryear{Matese \& O'Connell}{Matese \&
		O'Connell}{1969}]{Matese:1969cj}
	Matese J.~J.,  O'Connell R.~F.,  1969, \mn@doi [Phys. Rev.]
	{10.1103/PhysRev.180.1289}, 180, 1289
	
	\bibitem[\protect\citeauthoryear{Mesinger \& Furlanetto}{Mesinger \&
		Furlanetto}{2007}]{Mesinger:2007S}
	Mesinger A.,  Furlanetto S.,  2007, \mn@doi [ApJ] {10.1086/521806}, 669, 663
	
	\bibitem[\protect\citeauthoryear{Mesinger, Furlanetto  \& Cen}{Mesinger
		et~al.}{2011}]{Mesinger:2011FS}
	Mesinger A.,  Furlanetto S.,   Cen R.,  2011, \mn@doi [MNRAS]
	{10.1111/j.1365-2966.2010.17731.x}, 411, 955
	
	\bibitem[\protect\citeauthoryear{Mesinger, Ferrara  \& Spiegel}{Mesinger
		et~al.}{2013}]{Mesinger2013a}
	Mesinger A.,  Ferrara A.,   Spiegel D.~S.,  2013, \mn@doi [MNRAS]
	{10.1093/mnras/stt198}, 431, 621
	
	\bibitem[\protect\citeauthoryear{Minoda, Tashiro  \& Takahashi}{Minoda
		et~al.}{2019}]{Minoda:2018gxj}
	Minoda T.,  Tashiro H.,   Takahashi T.,  2019, \mn@doi [MNRAS]
	{10.1093/mnras/stz1860}, 488, 2001
	
	\bibitem[\protect\citeauthoryear{Mirocha \& Furlanetto}{Mirocha \&
		Furlanetto}{2019}]{Mirocha2019}
	Mirocha J.,  Furlanetto S.~R.,  2019, \mn@doi [MNRAS] {10.1093/mnras/sty3260},
	483, 1980
	
	\bibitem[\protect\citeauthoryear{Moroi, Nakayama  \& Tang}{Moroi
		et~al.}{2018}]{Moroi2018}
	Moroi T.,  Nakayama K.,   Tang Y.,  2018, \mn@doi [PLB]
	{10.1016/J.PHYSLETB.2018.07.002}, 783, 301
	
	\bibitem[\protect\citeauthoryear{Mu{\~{n}}oz \& Loeb}{Mu{\~{n}}oz \&
		Loeb}{2018}]{Munoz2018}
	Mu{\~{n}}oz J.~B.,  Loeb A.,  2018, \mn@doi [Nature]
	{10.1038/s41586-018-0151-x}, 557, 684
	
	\bibitem[\protect\citeauthoryear{Mu{\~{n}}oz, Dvorkin  \& Loeb}{Mu{\~{n}}oz
		et~al.}{2018}]{Munoz2018a}
	Mu{\~{n}}oz J.~B.,  Dvorkin C.,   Loeb A.,  2018, \mn@doi [PRL]
	{10.1103/PhysRevLett.121.121301}, 121, 121301
	
	\bibitem[\protect\citeauthoryear{Neronov \& Vovk}{Neronov \&
		Vovk}{2010}]{Neronov:1900zz}
	Neronov A.,  Vovk I.,  2010, \mn@doi [Science] {10.1126/science.1184192}, 328,
	73
	
	\bibitem[\protect\citeauthoryear{Olesen}{Olesen}{1997}]{OLESEN1997321}
	Olesen P.,  1997, \mn@doi [Physics Letters B]
	{https://doi.org/10.1016/S0370-2693(97)00235-9}, 398, 321
	
	\bibitem[\protect\citeauthoryear{Park, Mesinger, Greig  \& Gillet}{Park
		et~al.}{2019}]{Park:2019}
	Park J.,  Mesinger A.,  Greig B.,   Gillet N.,  2019, \mn@doi [MNRAS]
	{10.1093/mnras/stz032}, 484, 933
	
	\bibitem[\protect\citeauthoryear{Pavlovi\ifmmode~\acute{c}\else \'{c}\fi{},
		Leite  \& Sigl}{Pavlovi\ifmmode~\acute{c}\else \'{c}\fi{}
		et~al.}{2017}]{PhysRevD.96.023504}
	Pavlovi\ifmmode~\acute{c}\else \'{c}\fi{} P.,  Leite N.,   Sigl G.,  2017,
	\mn@doi [Phys. Rev. D] {10.1103/PhysRevD.96.023504}, 96, 023504
	
	\bibitem[\protect\citeauthoryear{{Planck Collaboration}}{{Planck
			Collaboration}}{2018}]{Planck:2018}
	{Planck Collaboration} 2018, \href
	{https://ui.adsabs.harvard.edu/abs/2018arXiv180706209P} {arXiv:1807.06209}
	
	\bibitem[\protect\citeauthoryear{{Planck Collaboration} et~al.,}{{Planck
			Collaboration} et~al.}{2014}]{2014A&A...571A..16P}
	{Planck Collaboration} et~al., 2014, \mn@doi [\aap]
	{10.1051/0004-6361/201321591}, 571, A16
	
	\bibitem[\protect\citeauthoryear{Pritchard \& Loeb}{Pritchard \&
		Loeb}{2012}]{Pritchard_2012}
	Pritchard J.~R.,  Loeb A.,  2012, \mn@doi [Rep. Prog. Phys]
	{10.1088/0034-4885/75/8/086901}, 75, 086901
	
	\bibitem[\protect\citeauthoryear{Schleicher, Banerjee  \& Klessen}{Schleicher
		et~al.}{2008}]{Schleicher:2008aa}
	Schleicher D. R.~G.,  Banerjee R.,   Klessen R.~S.,  2008, \mn@doi [Phys. Rev.
	D] {10.1103/PhysRevD.78.083005}, 78, 083005
	
	\bibitem[\protect\citeauthoryear{Seager, Sasselov  \& Scott}{Seager
		et~al.}{1999}]{Seager1999}
	Seager S.,  Sasselov D.~D.,   Scott D.,  1999, \mn@doi [Ast. J.]
	{10.1086/312250}, 523, L1
	
	\bibitem[\protect\citeauthoryear{Seager, Sasselov  \& Scott}{Seager
		et~al.}{2000}]{Seager}
	Seager S.,  Sasselov D.~D.,   Scott D.,  2000, \mn@doi [ApJ] {10.1086/313388},
	128, 407
	
	\bibitem[\protect\citeauthoryear{Sethi \& Subramanian}{Sethi \&
		Subramanian}{2005}]{Sethi:2004pe}
	Sethi S.~K.,  Subramanian K.,  2005, \mn@doi [MNRAS]
	{10.1111/j.1365-2966.2004.08520.x}, 356, 778
	
	\bibitem[\protect\citeauthoryear{Sethi, Nath  \& Subramanian}{Sethi
		et~al.}{2008}]{Sethi:2008eq}
	Sethi S.~K.,  Nath B.~B.,   Subramanian K.,  2008, \mn@doi [MNRAS]
	{10.1111/j.1365-2966.2008.13302.x}, 387, 1589
	
	\bibitem[\protect\citeauthoryear{Sharma}{Sharma}{2018}]{Sharma2018}
	Sharma P.,  2018, \mn@doi [MNRAS Lett.] {10.1093/mnrasl/sly147}, 481, L6
	
	\bibitem[\protect\citeauthoryear{Shu}{Shu}{1992}]{Shu:1992fh}
	Shu F.~H.,  1992, {The physics of astrophysics. Volume II: Gas dynamics.}.
	ISBN 0-935702-65-2, \url {http://adsabs.harvard.edu/abs/1992pavi.book.....S}
	
	\bibitem[\protect\citeauthoryear{Sikivie}{Sikivie}{2019}]{SIKIVIE2019100289}
	Sikivie P.,  2019, \mn@doi [Physics of the Dark Universe]
	{10.1016/j.dark.2019.100289}, 24, 100289
	
	\bibitem[\protect\citeauthoryear{Slatyer \& Wu}{Slatyer \&
		Wu}{2018}]{Slatyer2018}
	Slatyer T.~R.,  Wu C.-L.,  2018, \mn@doi [PRD] {10.1103/PhysRevD.98.023013},
	98, 023013
	
	\bibitem[\protect\citeauthoryear{Sur, Brandenburg  \& Subramanian}{Sur
		et~al.}{2008}]{Sur2008}
	Sur S.,  Brandenburg A.,   Subramanian K.,  2008, \mn@doi [MNRAS Lett.]
	{10.1111/j.1745-3933.2008.00423.x}, 385, L15
	
	\bibitem[\protect\citeauthoryear{Tashiro \& Sugiyama}{Tashiro \&
		Sugiyama}{2006}]{Tashiro:2005ua}
	Tashiro H.,  Sugiyama N.,  2006, \mn@doi [MNRAS]
	{10.1111/j.1365-2966.2006.10178.x}, 368, 965
	
	\bibitem[\protect\citeauthoryear{Tashiro, Kadota  \& Silk}{Tashiro
		et~al.}{2014}]{Tashiro:2014tsa}
	Tashiro H.,  Kadota K.,   Silk J.,  2014, \mn@doi [PRD]
	{10.1103/PhysRevD.90.083522}, 90, 083522
	
	\bibitem[\protect\citeauthoryear{Tavecchio, Ghisellini, Foschini, Bonnoli,
		Ghirlanda  \& Coppi}{Tavecchio et~al.}{2010}]{Tavecchio:2010GFB}
	Tavecchio F.,  Ghisellini G.,  Foschini L.,  Bonnoli G.,  Ghirlanda G.,   Coppi
	P.,  2010, \mn@doi [MNRAS: Letters] {10.1111/j.1745-3933.2010.00884.x}, 406,
	L70
	
	\bibitem[\protect\citeauthoryear{{The FLAT Collaboration} \& Biteau}{{The FLAT
			Collaboration} \& Biteau}{2018}]{Fermi_LAT:2018AB}
	{The FLAT Collaboration} Biteau J.,  2018, \mn@doi [ApJS]
	{10.3847/1538-4365/aacdf7}, 237, 32
	
	\bibitem[\protect\citeauthoryear{Trivedi, Seshadri  \& Subramanian}{Trivedi
		et~al.}{2012}]{Trivedi:2012ssp}
	Trivedi P.,  Seshadri T.~R.,   Subramanian K.,  2012, \mn@doi [PRL]
	{10.1103/PhysRevLett.108.231301}, 108, 231301
	
	\bibitem[\protect\citeauthoryear{Venumadhav, Dai, Kaurov  \&
		Zaldarriaga}{Venumadhav et~al.}{2018}]{PhysRevD.98.103513}
	Venumadhav T.,  Dai L.,  Kaurov A.,   Zaldarriaga M.,  2018, \mn@doi [PRD]
	{10.1103/PhysRevD.98.103513}, 98, 103513
	
	\bibitem[\protect\citeauthoryear{Wouthuysen}{Wouthuysen}{1952}]{1952AJ.....57R..31W}
	Wouthuysen S.~A.,  1952, \mn@doi [apj] {10.1086/106661}, 57, 31
	
	\bibitem[\protect\citeauthoryear{Yamamoto}{Yamamoto}{2016}]{Yamamoto:2016}
	Yamamoto N.,  2016, \mn@doi [Phys. Rev. D] {10.1103/PhysRevD.93.065017}, 93,
	065017
	
	\makeatother
\end{thebibliography}

\bsp	
\label{lastpage}
\end{document}